# Network Ecology of Marriage

Tamas David-Barrett

Email: tamas.david-barrett@trinity.ox.ac.uk

Address: Trinity College, Broad Street, Oxford, OX1 3BH, UK

Web: www.tamasdavidbarrett.com

**Abstract**

The practice of marriage is an understudied phenomenon in behavioural sciences despite being ubiquitous across human cultures. This modelling paper shows that replacing distant direct kin with in-laws increases the interconnectedness of the family social network graph, which allows more cooperative and larger groups. In this framing, marriage can be seen as a social technology that reduces free-riding within collaborative group. This approach offers a solution to the puzzle of why our species has this particular form of regulating mating behaviour, uniquely among pair-bonded animals.

# Introduction

Marriage is an enigma.

On one hand, it is a human near universal in the sense that it exists in almost every culture (1-6). On the other hand, its function is not clear.

Let us define 'marriage' as (i) a socially recognised agreement between at least two people such that (ii) they declare a romantic bond, which (iii) has an implied sexual component, with (iv) at least some barrier to separation. Despite it interacting with one of the most important biological behaviours, i.e., reproduction (7), the fact that marriage practices exist is relatively understudied by evolutionary behavioural science (8, 9). The literature's focus tends to be on the variation of the practice: for instance, in the number and sex of partners (10, 11), in partner selection (12), in within-marriage behavioural norms and resource flows (13, 14), and in inheritance (15, 16). In these papers, marriage is seen as a uniquely human regulatory system of mating, yet the reason for its existence is less clear. Why would marriage systems emerge in almost every human culture, given that no other animal, pair-bonded or not, has anything similar? What function of this practice is so useful that it is present all over our species' bewildering cultural kaleidoscope? And if there is such a super adaptive trait, why has it not appeared in any of the non-human animals? All animals that are mated for life manage to do so without the trappings of marriage: no siamang, owl monkey, swan, or bold eagle needs a socially enforced agreement to stay together. And no prairie vole, wolf, or cockatiel parrot is engaging in divorce procedure during relationship dissolution or mate switching.

The literature's answer to this is limited. The evolutionary origin of marriage is hazy (17), and so is its prevalence among human ancestors (18). It is clear that most marriage systems focus on a couple that is romantically bonded. The evolutionary origins of human pair bond have been widely theorised, but without consensus (19-24). However, it has, presumably, an ecological logic for our species, too, just as in the case of all other pair bonded animals. In any case, the establishment of a romantic bond during the infatuation phase is universal among humans (25), and is driven by a genetically inherited, mostly oxytocin-based bonding mechanism (26). In this, humans are far from unique (27).

All human culture's social organisation is multi-male, multi-female within which pair bonds emerge, a combination that is unique among primates (28). Although this setup allows substantial flexibility between bonobo-like human matriarchy (29) and chimpanzee-like human patriarchies (30), the two-way establishment of the pair-bond, popularly known as romantic infatuation, is an evolved, genetically inherited behaviour (31). For instance, both in partible paternity cultures (32) and modern polyamory cultures (33, 34) primary partners tend to be recognised as the main pair-bond, despite the frequent presence of other romantic partners.

Thus, although it is uncontroversial that all human cultures have some form of pair bond, and that this is an evolved behaviour in our species, none of these insights give a reason why the practice of marriage is so ubiquitous. In fact, the presence of a genetically inherited mechanism for establishing a human couple raises the question of why there is any need for the practice of marriage to start with.

It has been long established that marriage allows the extension of a cooperating group by including affinal kin, i.e., in-laws (35-37). However, this observation does not explain why a larger group is achieved via incorporating in-laws rather than inviting in somewhat more distant direct kin. For instance, while reciprocal exogamy does facilitate the presence of strong bond between two kin groups, it is not obvious why such a strong bond is needed to start with. If a group of people, bonded along direct kinship lines, need more people for their group, say for the technology they use would be more efficient with a large number of people, why don't they simply bring in more direct relatives instead of the affines?

The presence of marriage practices is perhaps even further puzzling given that although this social technology has been 'invented' by many cultures distant to each other, presumably independently, the past decades have brought a steady decline in marriage rates across the world (38, 39).

A recently proposed model system, the Structural Microfoundations Theory (40, 41), or Microfoundations for short, has directed focus to the importance of structural properties of social network in generating collective action (42). In the framework of the Microfoundations theory, kinship accounting is a heuristic that allows increased social network interconnectedness (40). This raises the question whether, perhaps, extending kin recognition to in-laws might also be about managing internal cooperativeness of a group. This yields the hypotheses of this paper.

H1: Affinal kinship increases the interconnectedness of a cooperating group.

If this prediction holds, then the institution of marriage could be seen as a social technology that reduces free riding by affecting the structure of the social network. If so, then the Microfoundations' proposed connection between demographic variables that work for kinship accounting would also work for creating in-laws via the marriage practice. Hence, a further hypothesis can be made:

H2: Falling fertility diminishes the importance of affinal kinship.

## Model

The model used the same populations as generated by the population histories described in the mathematical section of the paper 'Kinship is a Network Tracking Social Technology, Not an Evolutionary Phenomenon' (40). As it is described in that paper, the simulations grew populations through many generations, with parental female-male pairs assigned randomly, but not with sibling or first-degree cousins. All the 400 simulations used identical setup, except for the level of fertility in the society, which varied between 2.5 and 5 children per woman across the simulations and was uniform within each simulation across generations. In each simulation's population history, the final generation was selected as the focus population, when the simulation reached 2000 individuals. (For detailed description of how these population histories were generated, see 40). This method yielded 400 populations, each made up of 2000 same generation individuals, with their parents, grandparent, great-grandparents, and great-great grandparents recorded. This allowed the construction of a kinship network graph for each population the following way.

Let $M_g \equiv \{m_{i,j}\}$ denote the distance matrix of the kinship graph $g$, such that the element $m_{i,j}$ is the number of edges between vertices (i.e., individuals) $i$ and $j$ within the kinship tree. For instance, the distance between two siblings is 2, first degree cousins is 4, second degree cousins is 6. (For a detailed description of how to build a graph of kinship relationships, see 40.) The $M_g$ matrix's size is 2000x2000.

Let us assume, without further formal elaboration, that:

- agents form groups to perform collective action in which the individual contribution is costly, but gossip increases the cost of free-riding;
- gossip runs on social network triads, and thus groups with higher clustering coefficient are more cooperative (in line with 42);
- the cost of initiating and maintaining social network edges increases with network distance;

- and thus, that agents aim to form groups in which the average network path length is the shortest possible.

In line with these assumptions, I used the following algorithm to generate groups of agents:

Step 1. A random individual was chosen, denoted as $x_1$.

Step 2. Out of the set of individuals that were nearest to this individual (typically siblings), a second individual was chosen, denoted as $x_2$. Formally:

$$x_2 \sim U\{j | m_{x_1,j} = \text{Min}\{m_{x_1,i}\}_{i=1}^n, i \neq x_1\}$$

I added this second individual to the first to form the initial group as the $\{x_1, x_2\}$ set.

Step 3. Out of all the remaining individuals, the set of those individuals that were the nearest to the first two elements of the group was selected. A random element of this set was added to the group. Formally:

$$x_3 \sim U\{j | m_{x_1,j} = \text{Min}\{m_{x_1,i} + m_{x_2,i}\}_{i=1}^n, i \neq x_1\}$$

The algorithm, repeated step 3 until the group size reached 200, always choosing the next group member randomly among those that were the nearest on average to the already included group members.

**Fertility drives group integration.** For each of the 400 simulated populations, I generated 50 random groups using the above algorithm. I then calculated the average path lengths within the groups, as a function of the fertility. As a direct consequence of the way the groups were generated, i.e., that the next group member was always chosen as one that was the nearest to the already included group members, the average path lengths monotonously increase as a function of the group size (Fig. 1a).

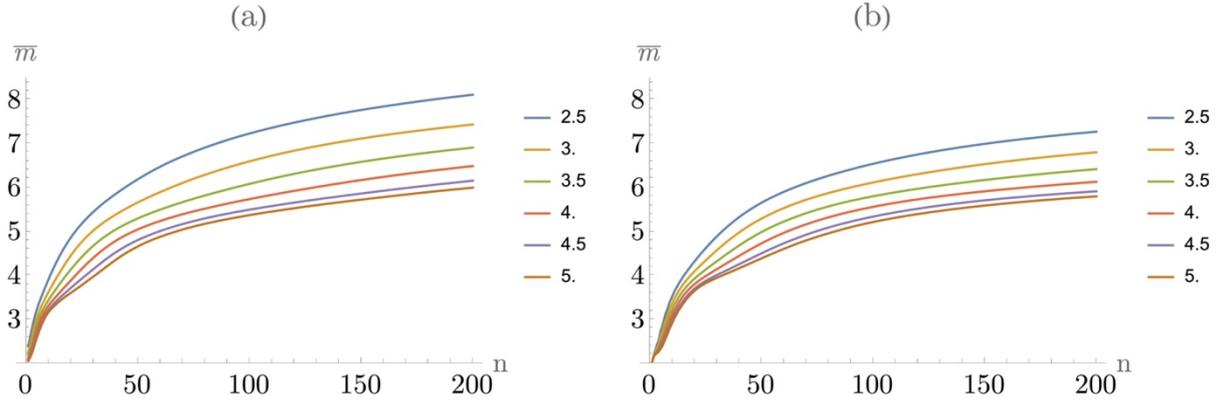

Fig. 1. Group size, fertility, and average path lengths. Panel (a): direct kin only; panel (b): both direct kin and affines. The average path length is increasing with group size, and decreasing with fertility for both the direct-kin-only and the combined direct-and-affine-kin networks. (The group size is denoted by $n$, the average path length within the group by $\bar{m}$.)

Notice that the population's fertility affects the average graph distance within the groups. As the fertility of the population drives the family size, it also drives the number of available same generation relatives. Thus, the higher the fertility, the smaller is the average path length. (NB. In this model, there is no mortality to reduce the effective family size, nor is there urbanisation or migration to make family members unavailable.)

The above model only recognises direct kin. To allow affinal kinship, I altered the kin graph, $g$, such that everyone is paired up with a random other individual as 'spouse'. This meant creating 1000 random pairs out of the 2000 individuals of the population and generating a network edge between each pair. At this point, the recognition of affines is possible. Assuming that the graph distance between 'spouses, is 1, I calculated the $M_g$ distance matrix for the new, affine-integrated graph, and built 50 random groups following the above steps 1-3 for each of the 400 populations. Panel (b) in Fig. 1 shows the resulting relationship between group size, fertility, and average path length. The results show that, similar to the direct kin-only graph, the path lengths monotonously increase as the group size increases, and that falling fertility increases the average path length.

Notice that the path length in the kin-and-affine graphs, i.e., panel (b), are shorter than their respective counterparts in the kin-only graphs, i.e., panel (a). That is, affinal kinship allows the creation of groups in which the average graph distance is shorter.

The pattern in the difference between the direct kin-only and the combined affinal and direct kin cases, i.e., Fig. 1a- Fig 1b, gives a clue about how this advantage happens (Fig. 2a).

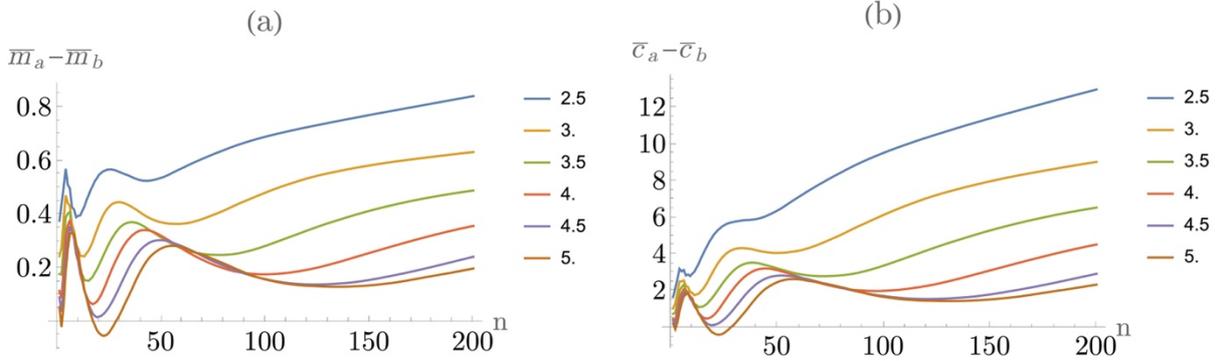

Fig. 2. Including affinal kin edges in the kinship graph reduces the average path length, and the group's cost of maintaining cohesion. Panel (a): the reduction in average path length due to recognising affinal edges, i.e., the difference Fig. 1a and 1b. Panel (b): the corresponding cost gain, assuming that the edge maintenance costs are quadratic. (The group size is denoted by $n$, the average path length within the group by $\bar{m}_a$ and $\bar{m}_b$ for the direct-kin-only, and the combined direct-and-affine-kin networks.

NB. The lowest curve in Fig. 2a, associated with fertility being 5, crosses the x-axis. Negative values are possible because the group formation algorithm is random among vertices that have the same average distance to the current group members. However, some of these will have different further vertex choice consequences later in the group build. (Computationally it would take way too long to calculate the cumulative future effect from each vertex and choose that way. In any case, in real-life, it is unlikely that people would think that strategically in their choices of new group members when extending a group.)

Furthermore, if the graph distance drives the cost of relationship bond, this result also means that including affinal edges lowers the cost of the entire social maintenance cost. To see this, let us assume that it is costly to maintain social network edges, and that this cost increases quadratically, i.e., $c = m^2$. The cost advantage reflects the average path length advantage: the social maintenance cost of group cohesion decreases when affinal social network edges are recognised (Fig. 2b).

## Direct and indirect affinal kinships differ in network properties

Notice the 'wavy' pattern Fig. 2. The first wave from the left is due to adding the immediate spousal edges of the group members. The second and third waves, however, are only possible if further away in-laws come into the group. Thus, for the maintained positive effect to be there, depicted in Fig. 2, the key is not only the relationship between an individual to his/her in-laws, but the relationship *among* the in-laws. It is easy to see why: as the distance between siblings is 2 and spouses 1, the distance between two in-law siblings, i.e., between the sibling of one spouse and the sibling of the other spouse, is 5. Thus, these affinal connections will be added to the group before the second-degree cousins are added (where the distance is 6).

To demonstrate this, I calculated the proportion of relationship types the following way.

Given a group, a random group member, $x1$, was selected. Then the shortest paths were identified between $x1$ and all the other group members. Each shortest path was categorised the following way. If the shortest path between $x1$ and $xi$ included only direct kin: '0'. If there is exactly one spousal edge in the shortest path, and it is at one of the ends of the path: 'a', and if it is in-between, i.e., not at the ends: 'b'. If there are two spousal edges, and both are at the shortest path's ends: 'aa'. If one is at the end, while the other is in-between: 'ab'. If both are in-between: 'bb'.

This categorisation allows a proximate approach to the importance of the direct kin ('0'), affinal kin via direct spousal connection ('a'+'aa'), and affinal kin via indirect spousal connection ('b'+'ab'+'bb') via measuring the number of these relationships inside the group.

The results offer two insights. First, depicted in Fig. 3, the aggregate shows that 1/2 to 2/3 of the group members are linked to each other where at least part of the shortest path includes one or more affinal edges, independent of group size and fertility. Incidentally, this is similar ratio that is present in hunter-gatherer societies (43). Notice also that the affines gain presence in middle sized groups compared to small and large, and increasing fertility somewhat decreases the importance of affinal kin compared to direct kin in large groups, although still above 50%.

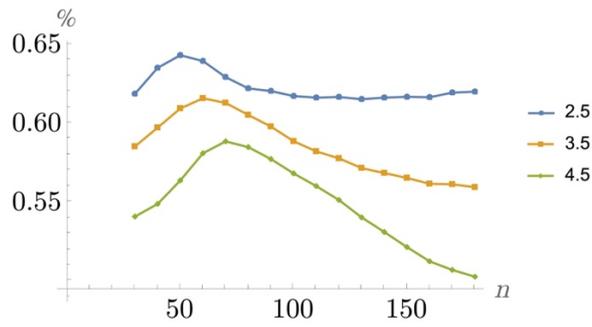

Fig. 3. The proportion of affinal kin relationships is between 50 and 65% for fertility 2.5-4.5, with the exact pattern affected by both fertility and group size.

Second, comparing the importance of direct spousal connections ('a'+'aa') to affinal kin via indirect spousal connection ('b'+'ab'+'bb'), the results show that the direct connection is important in all group sizes and for all fertility levels (left panel in Fig. 4). As group size increases the importance of the direct spousal link somewhat decreases for all fertility level, but especially for high fertility populations.

The pattern is exact opposite for indirect affinal kin: the overall importance (proportion) is relatively low but increases as the group size increases (right panel in Fig. 4).

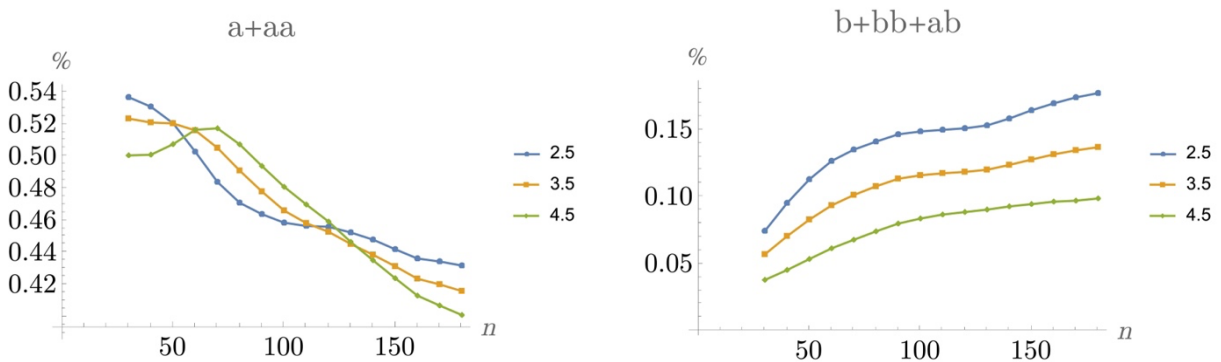

Fig. 4. Affinal kin relationship type composition for an average group member: as group size increases, the relative importance of direct affinal kin decreases, and of indirect affinal kin increases. (See Fig. S1 for all six individual types.)

The two affinal kin edge types are different in how their importance responds to increasing group size: when the affinal edge is at one or both ends of the shortest path, the importance of these is decreasing as the group size increases. At the same time, the in-between affinal edge is increasing in importance as the group size increases. (NB. This is true for 'ab' case, as well. For the detailed result affinal type-by-type ratios, see Fig. S1 in the Supplementary Material.)

### Minimum group integration threshold

Notice that if there is a limit to the graph distance among group members, the group size hits an upper threshold. Such a graph distance limit may be present if cooperation, which is the aim of forming the group in this model to start with, requires that the social network is integrated, i.e., the clustering cooperation is high (42). Thus, a minimum threshold of cooperativeness translates to a maximum average graph distance among the group members, which thus limits the maximum group size. Notice that this informs an evolutionary conundrum of our species concerning 'unnaturally' large group size (44), suggesting that even if kinship can be regarded as a means to achieve larger cooperative groups, this 'social technology' has an upper limit.

To see how such group size limit emerges, let us define such a limit as a threshold in the average path length inside a group. (Such a threshold would be a horizontal line in Fig. 1 cutting across the curves.) Notice that limiting the path lengths among the group members limits the size of the group that can be achieved, and thus yields a maximum group size variable. The maximum group size increases with fertility and the path-length threshold both in the direct kin-only graphs (Fig. 5a) and the graphs of combined direct and affinal kin (Fig. 5b).

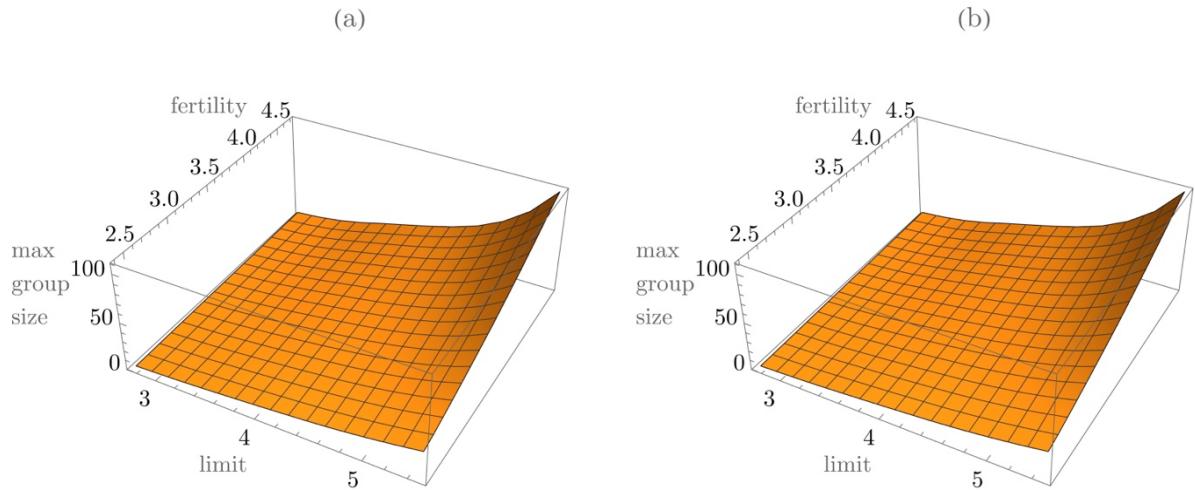

Fig. 5. Maximum group size increases with both fertility and the limit to average path length inside the group. Panel (a): direct kin only. Panel (b): both direct and affinal kin.

The difference between the two graph types (Fig. 6) shows the gain from including spousal links in kinship accounting. The difference between maximum group size in the combined affinal and direct kin and direct-kin only cases is always positive, independent of the average path length threshold or fertility level (Fig. 6a). In other words, the advantage from affinal kinship is that it allows larger groups at the same level of integration.

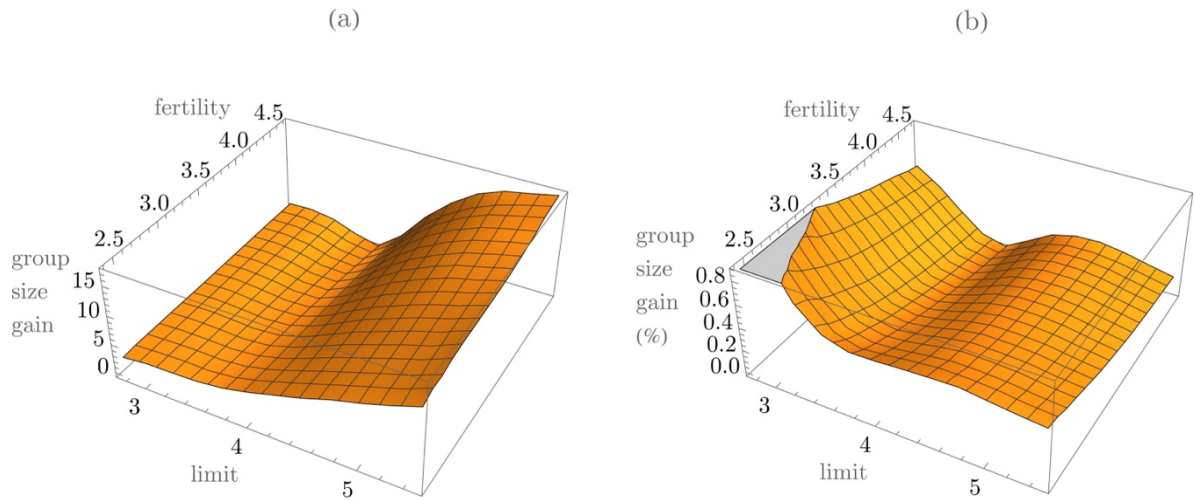

Fig. 6. Recognising affinal kin increases the maximum group size, given fertility and fixed limit to average path length. Panel (a): absolute gain, i.e., the difference between Fig. 5b and 5a. Panel (b): gain in percentage.

The proportional gain in maximum group size shows a generally downward slopping wavy pattern as a function of the limit, with a local minimum between path lengths 3 and 4 (Fig. 6b). This is due to the advantage that the sibling-spouse1-spouse2-sibling, which is three-edge length, affords the shortest type "b" relationship kind. Thus, this 'hump' is another illustration of the importance of in-between, indirect affinal relationships in kinship-based groups. (NB. While at the shortest limit, 3 and under, and especially at low fertility, the proportional gain is very high (the left corner of Fig. 6b), this corresponds to a low number of people (same corner of Fig. 6a).)

Notice that the path length limit plays a key role. This is especially important, as the Microfoundations models suggest a network trade-off between kinship and friendship (40, 45), such that as fertility falls, and hence as family ties give way to friendship ties, the kinship tracking path length decreases. For instance, when family size is large, the kinship tracking is likely to include every family member up to second degree cousin, but when family size is small, the tracking is reduced to first degree cousins, and then further to siblings only. If such a tracking limit reduction occurs, this would impact the tracking of in-laws, as well.

To see this, let us assume a negative linear relationship between fertility and the within-group average path lengths threshold, with a simple linear, negative correspondence between fertility ranging from 5 to 2.5, and the edge-path limit ranging from 5.5 to 2.75. (Notice that this is equivalent to drawing a straight line between the top right and bottom left corners in Fig. 5 and recording the curve's path.) Although the assumption of linearity is arbitrary, the negative correlation between fertility and path-length threshold is in line with the existing theoretical results (40, 41).

The results show that the gain from including affinal kin falls fast as fertility falls, with the difference from ~20 additional group members to ~2 (Fig. 7). (NB. The chosen limit range, i.e., 3-5 is merely for illustration purposes, if higher limit is allowed, for instance, if the collaborative task allows somewhat looser network integration, then the group size increases, and with that the absolute gain from tracking affines also increases.)

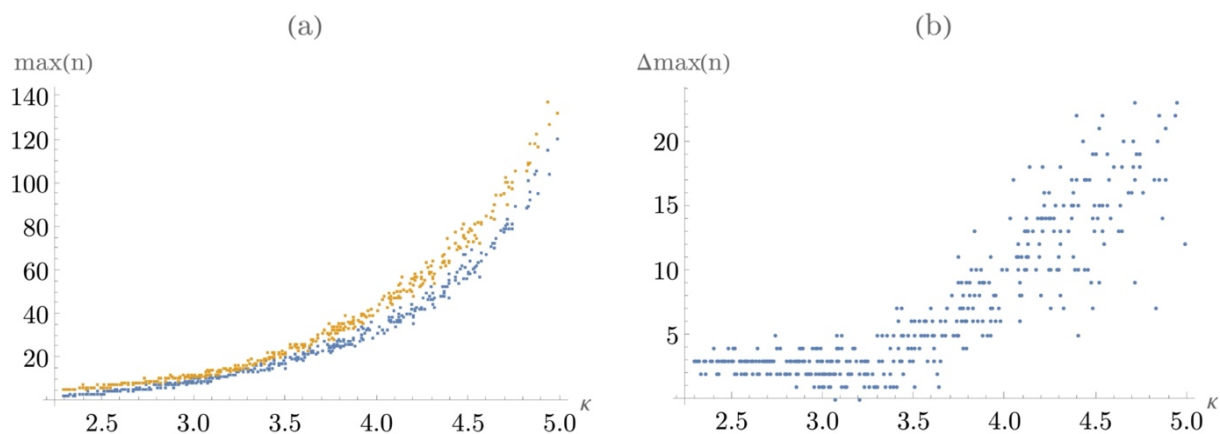

Fig. 7. Maximum group sizes, and group size gain from including affines, both as a function of fertility, given edge-length thresholds depending on fertility.

Notice this means that falling fertility leads to tracking fewer affines due to two different, parallel, effects. This is similar to tracking direct kin. First, falling fertility leads to fewer affines being around, given whatever path length limit is allowed by the nature of the collaborative act. Second, the affine-integrated social network's cooperativeness falls as the graph becomes less integrated, which leads to a reduced path length limit. As a consequence, falling fertility eliminates almost all gains from having affines instead of direct kin or friends, and thus makes the institution of marriage obsolete.

# Discussion

This paper aimed to create a formal framework in which the institution of traditional marriage is interpreted inside a network ecology.

The model presented in this paper suggests that tracking in-laws in a social network improves the cooperation of a mid-sized group. This can happen in two ways. If the minimum cooperative stance is fixed, then tracking in-laws increases the maximum size of the group. Alternatively (and equivalently), if the group size is fixed then extending kinship to affines, increases the cooperative stance of the group. In this sense, marriage can be regarded as a social technology that facilitates larger and/or more cooperative kin groups.

The model also showed that affines are similar to kin concerning the impact of falling fertility on the social network. When family size falls, and thus fewer direct and affinal kin are available, the importance of tracking network connections through marriage falls, to the point of negligible.

Furthermore, the model differentiated between two different kinds of affinal kinship: direct and indirect. The direct affinal kinship type is when the path between two individuals start, ends, or both in a married couple, and while there is no other married couple in the path. For instance, ego – spouse – spouse's parent – spouse's parent's other child is such a direct affinal kinship path. The indirect affinal kinship type is when there is at least one married couple on the path, but at neither end. For instance, ego – ego's parent – ego's parent's other child – ego's parent's child's spouse - ego's parent's child's spouse's parent - ego's parent's child's spouse's parent's other child. Notice that, in the English language affinal kinship tracking tradition, both of these are shortened as ego to sibling in-law relationship. The model showed that both of these two different kinds of affinal relationships are important, and their behaviour is somewhat different.

The findings of this paper yield a range of empirical predictions.

In this model's framing, formal marriage is a social technology that facilitates free-rider free cooperation in a relevant network ecology. This happens by marriage re-structuring the social network in a way that is important for organising collective action. We should expect that (a) people would take aspect of marriage into account when choosing a romantic partner, and across all cultures in which cooperation-enhancing logic of marriage is present, (b) the initiation of these cross-family bonds should be formalised, i.e., these cultures should invent the concept of a wedding, and (c) the dissolution of marriage should be regulated, especially that it should be made difficult.

**(a) Family compatibility considerations in mate choice.** When choosing a long-term romantic partner, people reorganise their social network, as described in this paper.

During mate choice, at the phase of assessing a potential spousal edge, it is important to 'check' not only if the edge itself will work between the two people who will form the new couple, but also if the new social network edges across the family will work well. We should expect to see people considering both direct and indirect affinal kin edges. In other words, when one meets a new potential partner, one should assess one's compatibility with the new partner's family, the compatibility of the potential new partner with one's own family, and the compatibility of the two families together (independent of the quality of the spousal edge itself).

That is, if I was to assess a potential new partner, I would need to check three different aspects. First, if my new partner and I will work well together, if we will fit. Much of the mate choice literature concerns this aspect (e.g., 46, 47). Second, I would need to check if my partner would get along with my family, and if I would get along with my partner's family. There is some literature on this, but limited to parental influence on mate choice (48-50). Third, I would need to check if my family will work well with my partner's family, i.e., if the two families are compatible such that the indirect affinal edges will work. As far as I know, this is the first paper that raises this.

**(b) Weddings are efficient polyadic edge creation events.** For the new, affinal social network edges to work, it is not enough to assess whether they would do so, but these edges also need to be operationalised. Therefore, we should expect that cultures in which marriage is practiced for the purpose of creating the cross-family bonds, should invent the social technology of an event in which the new affinal edges are established in an efficient way. In this sense, weddings can be seen as mass edge-creation events that use a range of bonding tricks, like sharing food, sharing supernatural narratives about imaginary agents, singing together, dancing together, playing together.

**(c) When affinal edges are important, divorce should be difficult.** The risk in structuring the collaborative social network via marriage is that be doing so, the group's collective action becomes dependent on a social network edge between two people, i.e., the married couple. Hence, spousal relationship dynamics can threaten the economic action of an entire group. We should expect that formal marriage institutions should regulate the factors that make dissolution more likely, for instance, romantic relationships outside the couple or the intimacy between the couple, and the severing of the edge should be made costly. That is, we should see that cultures that use this paper's logic of marriage make divorce less likely and difficult.

**Fertility drives marriage practice.** As the model of this paper showed, the importance of affinal edges is dependent on family size which in turn depends on fertility. In this sense, affinal edges work similar to direct kin edges as described by the Microfoundations theory (40-42, 45). This translates into the prediction that demographic transition should drive several aspects of the marriage: the weight of family compatibility during mate choice, the frequency of the marriage practice, the importance of weddings, and the regulation of divorce should all decrease as a consequence of falling fertility. Furthermore, we should expect a lagged effect between fertility transition and the importance of marriage as the number of adult relatives is dependent on the previous generations' fertility, and this delay should be generation length.

# References

1.      Marlowe FW. The mating system of foragers in the standard cross-cultural sample. Cross-Cultural Research. 2003;37(3):282-306.

2.      Lévi-Strauss C. Les structures élémentaires de la parenté. Paris: Presses universitaires de France; 1949. xiv, 639 p. p.

3.      Chapais B. The Deep Structure of Human Society: Primate Origins and Evolution. Mind the Gap: Tracing the Origins of Human Universals. 2010:19-51.

4.      Chapais B. Primeval kinship : how pair-bonding gave birth to human society. Cambridge, Mass. ; London: Harvard University Press; 2008. xv, 349 p. p.

5.      Walker RS, Hill KR, Flinn MV, Ellsworth RM. Evolutionary history of hunter-gatherer marriage practices. PloS one. 2011;6(4):e19066.

6.      Ember CR, Gonzalez B, McCloskey D. Marriage and Family: HRAF: Explaining Human Cultures; 2021.

7.      Westermarck E. The history of human marriage. London,: Macmillan; 1891. xix, 644 p. p.

8.     Low BS. Ecological and socio-cultural impacts on mating and marriage systems. In: Dunbar RIM, Barrett L, editors. Oxford handbook of evolutionary psychology. Oxford: Oxford University Press; 2007. p. 449–62.

9.     Fortunato L. Marriage Systems, Evolution of.  International Encyclopedia of the Social & Behavioral Sciences2015. p. 611-9.

10.    Reichard UH, Boesche C. Monogamy : mating strategies and partnerships in birds, humans and other mammals. Cambridge: Cambridge University Press; 2003. ix, 267 p. p.

11.    Starkweather KE, Hames R. A survey of non-classical polyandry. Human nature. 2012;23(2):149-72.

12.    Geary DC, Vigil J, Byrd-Craven J. Evolution of human mate choice. J Sex Res. 2004;41(1):27-42.

13.    Fortunato L, Holden C, Mace R. From bridewealth to dowry? A Bayesian estimation of ancestral states of marriage transfers in Indo-European groups. Hum Nature-Int Bios. 2006;17(4):355-76.

14.    Marlowe FW. Paternal investment and the human mating system. Behavioural processes. 2000;51(1-3):45-61.

15.    Fortunato L, Archetti M. Evolution of monogamous marriage by maximization of inclusive fitness. Journal of evolutionary biology. 2010;23(1):149-56.

16.    Goody J. The oriental, the ancient and the primitive : systems of marriage and the family in the pre-industrial societies of Eurasia. Cambridge: Cambridge University Press; 1990. xix, 542 p. p.

17.    Fortunato L. Reconstructing the history of marriage strategies in Indo-European-speaking societies: monogamy and polygyny. Human biology. 2011;83(1):87-105.

18.    Bittles AH, Black ML. Consanguineous Marriage and Human Evolution. Annual Review of Anthropology. 2010;39(1):193-207.

19.    Buss DM, Schmitt DP. Sexual Strategies Theory - an Evolutionary Perspective on Human Mating. Psychol Rev. 1993;100(2):204-32.

20.    Schacht R, Kramer KL. Are We Monogamous? A Review of the Evolution of Pair-Bonding in Humans and Its Contemporary Variation Cross-Culturally. Frontiers in Ecology and Evolution. 2019;7.

21. Fletcher GJ, Simpson JA, Campbell L, Overall NC. Pair-bonding, romantic love, and evolution: the curious case of Homo sapiens. Perspect Psychol Sci. 2015;10(1):20-36.

22. Gavrilets S. Human origins and the transition from promiscuity to pair-bonding. Proceedings of the National Academy of Sciences of the United States of America. 2012;109(25):9923-8.

23. Coxworth JE, Kim PS, McQueen JS, Hawkes K. Grandmothering life histories and human pair bonding. Proceedings of the National Academy of Sciences of the United States of America. 2015;112(38):11806-11.

24. Opie C, Atkinson QD, Dunbar RIM, Shultz S. Male infanticide leads to social monogamy in primates. Proceedings of the National Academy of Sciences of the United States of America. 2013;110(33):13328-32.

25. Jankowiak WR, Fischer EF. A Cross-Cultural Perspective on Romantic Love. Ethnology. 1992;31(2):149-55.

26. Walum H, Lichtenstein P, Neiderhiser JM, Reiss D, Ganiban JM, Spotts EL, et al. Variation in the oxytocin receptor gene is associated with pair-bonding and social behavior. Biol Psychiatry. 2012;71(5):419-26.

27. Young LJ, Wang Z. The neurobiology of pair bonding. Nature neuroscience. 2004;7(10):1048-54.

28. Shultz S, Opie C, Atkinson QD. Stepwise evolution of stable sociality in primates. Nature. 2011;479(7372):219-22.

29. Mattison SM. Demystifying the mosuo: The behavioral ecology of kinship and reproduction in china's "last matriarchal" society. ProQuest Dissertations & Theses Global: University of Washington; 2010.

30. Buss DM. Evolutionary psychology : the new science of the mind. 6th Edition. ed. New York: Routledge; 2019. xiv 503 pages p.

31. Diamond LM. Emerging perspectives on distinctions between romantic love and sexual desire. Current Directions in Psychological Science. 2004;13(3):116-9.

32. Walker RS, Flinn MV, Hill KR. Evolutionary history of partible paternity in lowland South America. Proceedings of the National Academy of Sciences of the United States of America. 2010;107(45):19195-200.

33. Balzarini RN, Campbell L, Kohut T, Holmes BM, Lehmiller JJ, Harman JJ, et al. Perceptions of primary and secondary relationships in polyamory. PloS one. 2017;12(5):e0177841.

34. Balzarini RN, Dharma C, Kohut T, Campbell L, Lehmiller JJ, Harman JJ, et al. Comparing Relationship Quality Across Different Types of Romantic Partners in Polyamorous and Monogamous Relationships. Archives of sexual behavior. 2019;48(6):1749-67.

35. Hughes AL. Evolution and human kinship. New York ; Oxford: Oxford University Press; 1988. viii, 162 p. p.

36. Burton-Chellew MN, Dunbar RIM. Are Affines Treated as Biological Kin? A Test of Hughes's Hypothesis. Current Anthropology. 2011;52(5):741-6.

37. Parkin R. Kinship : an introduction to basic concepts. Oxford: Blackwell; 1997. xi, 208 p. p.

38. Stevenson B, Wolfers J. Marriage and divorce: Changes and their driving forces. J Econ Perspect. 2007;21(2):27-52.

39. Posel D, Rudwick S. Changing patterns of marriage and cohabitation in South Africa. Acta Juridica. 2013;2013(1).

40. David-Barrett T. Kinship Is a Network Tracking Social Technology, Not an Evolutionary Phenomenon. arXiv. 2022(2203.02964).

41. David-Barrett T. Network Effects of Demographic Transition. Scientific reports. 2019;9.

42. David-Barrett T. Clustering drives cooperation on reputation networks, all else fixed. Roy Soc Open Sci. 2023;10(4).

43. Dyble M, Salali GD, Chaudhary N, Page A, Smith D, Thompson J, et al. Sex equality can explain the unique social structure of hunter-gatherer bands. Science. 2015;348(6236):796-8.

44. David-Barrett T. Human Group Size Puzzle: Why It Is Odd That We Live in Large Societies. bioRxiv. 2022(2022.02.18.481060).

45. David-Barrett T. Herding Friends in Similarity-Based Architecture of Social Networks. Scientific reports. 2020;10(1).

46. Buss DM. The evolution of human mating. Acta Psychologica Sinica. 2007:502–12.

47.	Walter KV, Conroy-Beam D, Buss DM, Asao K, Sorokowska A, Sorokowski P, et al. Sex Differences in Mate Preferences Across 45 Countries: A Large-Scale Replication. Psychological science. 2020;31(4):408-23.

48.	Apostolou M. Elements of parental choice: The evolution of parental preferences in relation to in-law selection. Evol Psychol-Us. 2007;5:70-83.

49.	Apostolou M. Parental choice: What parents want in a son-in-law and a daughter-in-law across 67 pre-industrial societies. Brit J Psychol. 2010;101:695-704.

50.	Apostolou M. Sexual Selection under Parental Choice: Evidence from Sixteen Historical Societies. Evol Psychol-Us. 2012;10(3):504-18.

# Supplementary Material

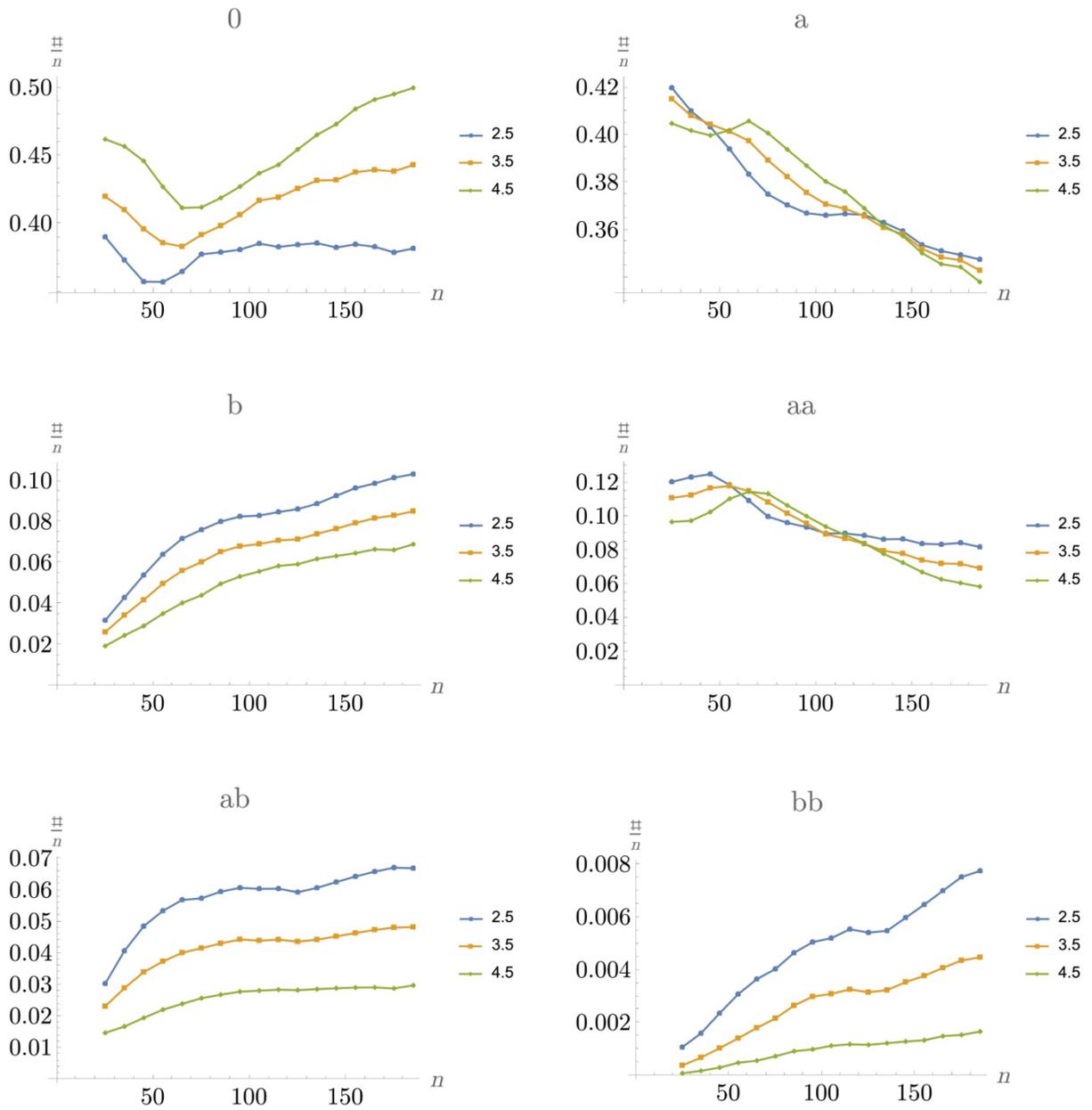

Fig. S1. An average group member's proportions of relationship types as a function of group size and fertility.